\documentclass[10pt]{cernrep_a}
\usepackage{graphicx}
\usepackage{floatflt}

\newcommand{\Pt}{{P_t}}
\newcommand{\dphi}{\Delta\phi}
\newcommand{\phigj}{\phi_{(\gamma,jet)}}

\newcommand{\la}{\langle}
\newcommand{\ra}{\rangle}
\newcommand{\gpj}{~"$\gamma+Jet$"~}
\newcommand{\rrr}{\to} 

\newcommand{\Db}{\Pt(O+\eta>5)}
\newcommand{\Ptg}{\Pt^{\gamma}}

\newcommand{\nuj}{Pt^{Jet}\! \! -\! \! Pt^{jet}}
\newcommand{\ptgj}{$~\Pt^{\gamma}$ and $\Pt^{jet}~$}

\suppressfloats[!]

\begin{document}
\thispagestyle{empty}

\vskip-5mm

\begin{center}
{\Large JOINT INSTITUTE FOR NUCLEAR RECEARCH}
\end{center}

\vskip10mm

\begin{flushright}
JINR Preprint \\
E2-2000-251 \\
hep-ex/0011012
\end{flushright}

\vspace*{3cm}

\begin{center}
\noindent
{\Large{\bfseries Jet energy scale setting with \gpj events at LHC energies.\\[5pt]
Generalities, selection rules.}}\\[5mm]
{\large D.V.~Bandourin$^{1\,\dag}$, V.F.~Konoplyanikov$^{2\,\ast}$, 
N.B.~Skachkov$^{3\,\dag}$}

\vskip 0mm

{\small
{\it
E-mail: (1) dmv@cv.jinr.ru, (2) kon@cv.jinr.ru, (3) skachkov@cv.jinr.ru}}\\[3mm]
$\dag$ \large \it Laboratory of Nuclear Problems \\
\hspace*{-4mm} $\ast$ \large \it Laboratory of Particle Physics
\end{center}

\vskip 9mm
\begin{center}
\begin{minipage}{150mm}
\centerline{\bf Abstract}
\noindent
\gpj events, based on the $q\bar{q}\to g+\gamma$ and $qg\to q+\gamma$
subprocesses, are proposed for jet energy scale setting and hadron calorimeter
calibration at LHC energies. General features and selection criteria of
\gpj events that would provide a good $\Pt^{\gamma}-\Pt^{Jet}$ balance
are described. CMS detector geometry is taken as the basement.
\end{minipage}
\end{center}

\newpage

\vskip2cm
\setcounter{page}{1}
\section{INTRODUCTION} 

Setting an absolute scale for a hadronic calorimeter
(HCAL) is an important task for
many of $pp$ and $p\bar{p}$ collider experiments (see e.g. [1, 2]).
 There is a number
of ATLAS and CMS publications on this subject where the application of
different physical processes (like "$Z^0/\gamma+Jet$" and others)
is discussed ([3--10]).

This paper is the first part of a set of publications on a detailed
study of this problem. The main goal
of this work is to find out the selection criteria for \gpj events that
would lead to the most precise determination of transverse momentum of a jet,
i.e. $\Pt^{Jet}$, via
assigning a photon $\Ptg$ to a jet. We shall present here
the results of event generation by using PYTHIA 5.7 [11].
Further development, based on the simulation of detector
response with GEANT detector simulation packages will be presented in
one of our next papers.

It should be also noted that here we consider the case of a low
luminosity $L=10^{33}\,cm^2 s^{-1}$ that is still quite sufficient
to use much more restrictive cuts as well as new physical variables
and, correspondingly, cuts on them
(cluster suppression criterion, for instance)
in comparison with those used in the previous experiments and to obtain
a set of more clean \gpj events.

The Section 2~
is an introduction to the problem.
General features of the \gpj processes at LHC energies, that will be
explored in this article and later papers, are presented here.
In Section 2 we review possible sources of the
$\Pt^{\gamma}$ and $\Pt^{Jet}$ disbalance in the final state and the ways of
selecting those events where this disbalance has a minimal value.

In Section 3.1 we present definitions of $\Pt$ components of
different objects that enter the balance equation  illustrating the
conservation law of the total $\Pt$ in any event.

Section 3.2 describes the criteria we have chosen to select \gpj events
for the calibration procedure.
"Cluster" (or mini--jet) suppression criterion ($\Pt^{clust}_{CUT}$) which
has not been used in previous experiments is introduced here. Its
important role will be illustrated in the following papers [12--15].
These clusters have a physical sense as a part of another new experimentally
measurable quantity
introduced here for the first time, namely, the sum of the $\vec{\Pt}$ of all
particles detectable in the $|\eta| < 5$ region which are out of the \gpj
system (denoted as $\Pt^{out}$).

Another new thing here is an introduction of a new physical object, named
as an ``isolated jet'',
i.e. the jet that does not have any noticeable $\Pt$ activity in some ring in
$\eta-\phi$ space around
it. In other words we will select some class of events
having  a total $\Pt$ activity
inside the ring, $\Delta R = 0.3$, around this ``isolated jet''
within $2-5\%$ of jet $\Pt$.
In the following paper it will be shown
that  the number of events
with such a clean topological structure would not be small at LHC
energies (mainly due to the growth of luminosity).

Since the calibration is rather a practical than an academical task
in all the following Sections, we present the values of rates for strict
and weak cut values because their choice would be
a matter of step-by-step collected statistics.

The justification of the variables
and cuts introduced in Section 3 can be found in our
papers [13--15]. In [15]
we present the estimation of the efficiency of background suppression,
that is, finally, the main guideline to establish the selection
rules.

Section 4 will be devoted to the estimation of
non-detectable neutrino contribution to $\Pt^{Jet}$
as well as to studying the influence of the $|\eta|>5$ region 
not covered by calorimeters or other detectors
(that is the main source of
$\Pt^{miss}\equiv \not\!\!{E_T}$) on the total $\Pt$ balance in the event. 
The correlation of the upper cut on $\Pt^{miss}$ with a mean value of $\Pt$ of neutrinos
belonging to the jet $\Pt$, i.e. $\la \Pt_{(\nu)}^{Jet}\ra$,
will be considered here.

Since the results presented here have been obtained with PYTHIA simulation,
we are planning to carry out the analogous estimation in the next papers  but
with another event generator.

\section{GENERALITIES OF \gpj PROCESSES}  
\vskip0.3cm
\subsection{Leading order picture}        

\setcounter{equation}{1}

The idea of a hadronic calorimeter (HCAL) calibration by physical
process "$pp\rrr \gamma+Jet$"
was realized many times in different experiments
(see recent papers [1, 2] and refs. there).
It is based on the parton picture where two partons ($q\bar{q}$ or
$qg$), supposed to be moving in different colliding nucleons with
zero transverse momenta
(with respect to the beam line), produce a photon, called a direct
one, and a parton with balanced transverse momentum
$\vec{\Pt}^{part}=-\vec{\Pt}^{\gamma}$.  This picture corresponds to
the leading order (LO) Feynman diagrams shown in Fig.~1
for the ''Compton-like'' process\\[-17pt]
\begin{eqnarray}
\hspace*{4.74cm} qg\to q+\gamma \hspace*{6cm} (1a)
\nonumber
\end{eqnarray}
\vspace{-3mm}
and for ''annihilation'' process\\[-10pt]
\begin{eqnarray}
\hspace*{4.62cm} q\overline{q}\to g+\gamma,  \hspace*{6cm} (1b)
\nonumber
\end{eqnarray}
respectively.
The $\Pt$ of the ``$\gamma$+parton'' system produced in the final state
should be equal to zero, i.e.\\[-10pt]
\begin{eqnarray}
\vec{\Pt}^{\gamma+part}=\vec{\Pt}^{\gamma}+\vec{\Pt}^{part} = 0.
\end{eqnarray}
So, in this case one
could expect that with a reasonable precision the transverse momentum
of the jet produced by the final state parton ($q$ or $g$) will
be close
in magnitude to the transverse momentum of the final state photon,
 i.e. $\vec{\Pt}^{Jet}\approx-\vec{\Pt}^{\gamma}$. \\[-6.2cm]
\begin{center}
\begin{figure}[h]
  \hspace{.15cm} \includegraphics[width=13cm,height=9cm]{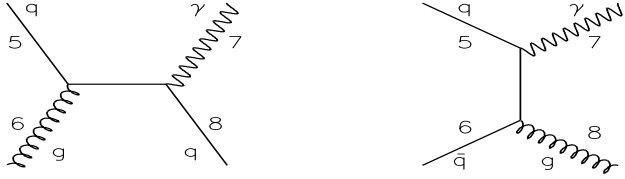} 
  \vspace{-18mm}
\caption{\hspace*{0.0cm} Some of the leading order Feynman diagrams for direct
photon production.} \label{fig1:LO}
\end{figure}
\end{center}
\vskip-0.9cm

It allows one to carry out the calibration of HCAL in the experiments
with a well calibrated electromagnetic calorimeter (ECAL). To put it simpler,
to a part of jet transverse energy $E_t^{Jet}$ deposited in HCAL we can assign
the value of the difference between the value of
the transverse energy deposited in ECAL in the photon direction,
 i.e. $E_t^{\gamma}$, and the value of the transverse energy deposited in ECAL in the jet direction.

\subsection{Initial state radiation.}                           

Since we believe in the perturbation theory, the leading
order (LO) picture, described above, is expected to be dominant in
determination of the main contribution to the
cross section. The Next-to-Leading Order
(NLO) approximation (see some of the NLO diagrams in Fig.~2) introduces some
deviations from a
rather straightforward LO-motivated idea of calibration. Thus, as it is seen from Fig.~2,
 ~a gluon ~radiated in
the initial state (ISR) can have its own non-zero transverse momentum
$\Pt^{ISR}\neq 0$. It leads
to the non-zero transverse momenta of partons that appear in the initial state
of fundamental $2\rrr2$ QCD subprocesses (1a) and (1b). As a result of
the transverse momentum conservation, a disbalance of the transverse momentum
of a photon $\Pt^{\gamma}$ and of a parton $\Pt^{part}$ produced in the fundamental
$2\to 2$ process $5+6\to 7+8$, shown in Figs.~2 and 3
(and thus, finally, of a jet produced by this parton), will take place.
\\[-6.3cm]
\begin{center} \begin{figure}[h]
  \hspace{-.5cm} \includegraphics[width=13cm,height=9cm,angle=0]{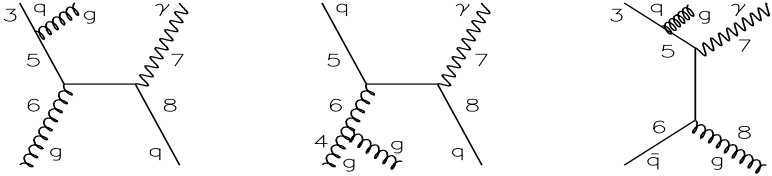}
  \vspace{-12mm}
  \caption{\hspace*{0.0cm} Some of Feynman diagrams of direct photon
production including gluon radiation in the initial state.}
    \label{fig2:NLO}
  \end{figure}
\end{center}
\vspace{-1.0cm}

We shall choose the modules of a vector sum of
transverse momentum vectors $\vec{\Pt}^{5}$ and $\vec{\Pt}^{6}$
of incoming  (into $2\rrr 2$ fundamental QCD subprocesses $5+6\to 7+8$) partons,
shown on lines 5 and 6 in Fig.~2,
as well as the sum of their modules
as two quantitative measures \\[-5pt]
\vspace{-3mm}
\begin{eqnarray}
\Pt^{5+6}=|\vec{\Pt}^5+\vec{\Pt}^6|, \qquad \Pt{56}=|\Pt^5|+|\Pt^6|
\end{eqnarray}
to estimate the $\Pt$ disbalance caused by ISR. The modules of the vector sum  \\[-5pt]
\vspace{-3mm}
\begin{eqnarray}
\Pt^{\gamma+Jet}=|\vec{\Pt}^{\gamma}+\vec{\Pt}^{Jet}| .
\end{eqnarray}
will be used as an estimator of the final state  $\Pt$
disbalance in the \gpj system.

The numeration notations in these Feynman diagrams as well as in formulae
(3) and (4) are chosen to be in correspondence with those
used in PYTHIA for describing parton--parton subprocess,
displayed schematically in Fig.~3. The ``ISR'' block describes the initial
state radiation process that can take place before the fundamental
hard $2\to 2$ process.
\begin{center}
  \begin{figure}[h]
  \vspace{-0.6cm}
   \hspace{1.8cm} \includegraphics[width=9cm,height=5cm,angle=0]{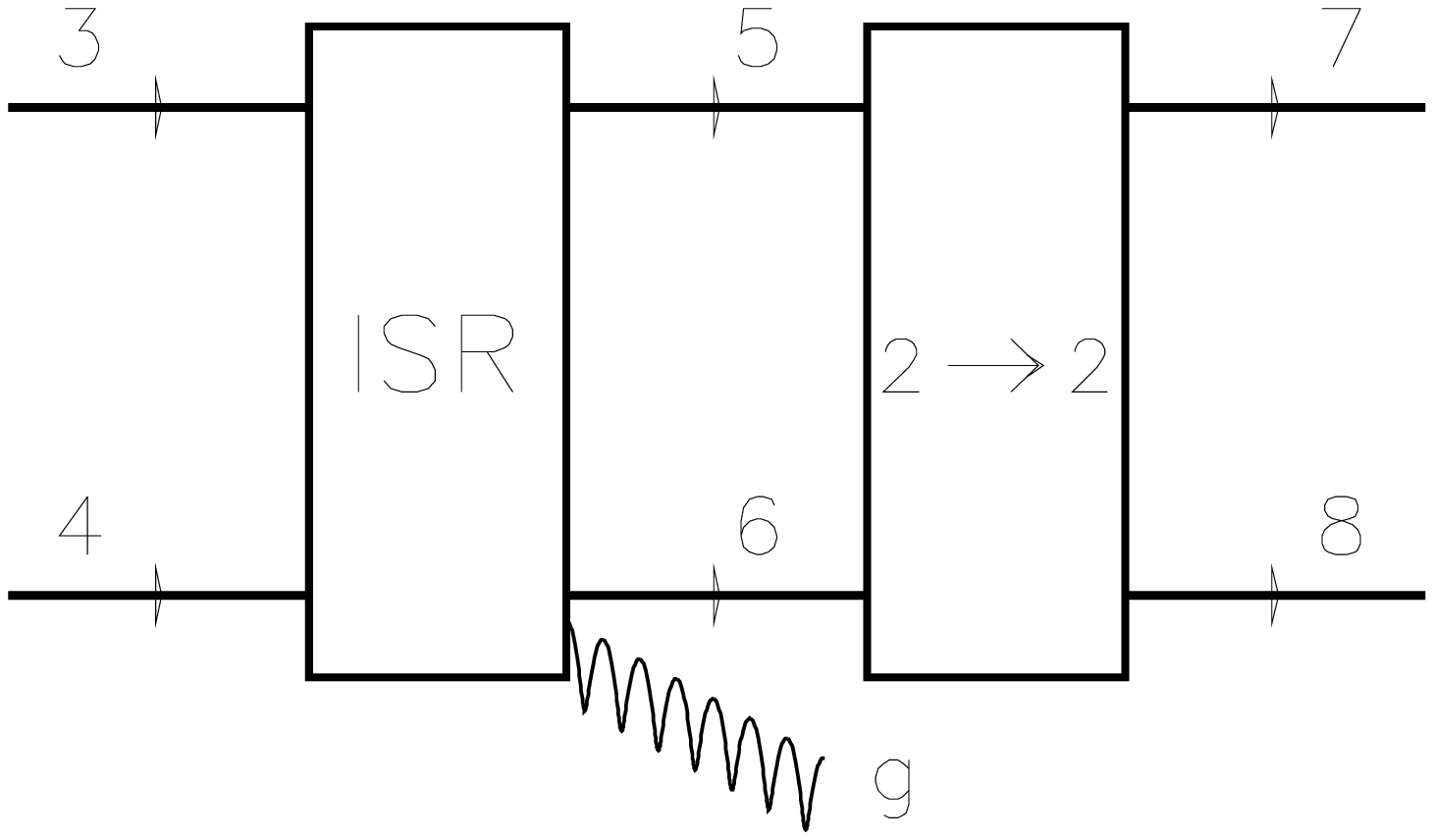}
  \vspace{-2.2cm}
 \caption{\hspace*{0.0cm} PYTHIA ``diagram'' of~ $2\to2$ process (5+6$\to$7+8)
with a block (3+4$\to$5+6) of initial state radiation (ISR).}
    \label{fig3:PYT}
  \end{figure}
\end{center}
  \vspace{-7mm}

\subsection{Primordial parton $k_T$ effect.}                   

A possible non-zero value of the intrinsic parton velocity inside a colliding proton
may be another source of the $\Pt^{\gamma}$ and $\Pt^{part}$
disbalance in the final state .
Its reasonable value is supposed to lead to the value of
$k_T \leq \,1.0 ~GeV/c$.
It should be noted that sometimes in the literature
the summarized effect of ISR and of the intrinsic parton transverse momentum
is denoted by $k_T$. Here we follow the approach used in PYTHIA where these
two sources of disbalance
are treated separately and switched on by different keys (MSTP(61) for ISR and
PARP(91), PARP(93) and MSTP(91) for $k_T$). Below we shall
keep the value of $k_T$ to be fixed  by PYTHIA  default value
$\langle k_T \rangle$=0.44 GeV/c. Its possible variation influence on
$\Pt^{\gamma}$ and $\Pt^{Jet}$ disbalance will be discussed in detail in our
following paper [15]. The general conclusion from there is that any variation
of $k_T$ within reasonable boundaries (as well as beyond them) does not produce
a large effect in the case when the initial state radiation is switched on. The last one
gives a dominant contribution.

\section{CHOICE OF MEASURABLE PHYSICAL VARIABLES FOR \gpj
PROCESS AND OF CUTS FOR BACKGROUND REDUCTION.}               
Another than (1a) and (1b) QCD processes with large cross sections, being
by orders of magnitude higher than the
\gpj cross section, can also contain high $\Pt$ photons
and jets in final states. So, we face the problem of
signal \gpj events selection out from the large QCD background.
Here we shall discuss the choice of physical variables that would be
useful under some cuts on their values to select the
desirable processes with direct photons (``$\gamma^{dir}$'') production
from the background events. The possible $\gamma^{dir}$ candidate
may originate from the $\pi-,~\eta-,~\omega-,~K-$ meson decays or, may
be, from a bremsstrahlung photon.

We suppose the ECAL size to be limited by
$|\eta| \leq 2.61$ and HCAL to consist of the barrel (HB), end-cap
(HE) and forward (HF) parts and to be limited by $|\eta| \leq 5.0$
(CMS geometry),
 where $\eta = -0.5\,ln~(tan~(\theta/2))$ is a pseudorapidity
defined through
a polar angle $\theta$ counted from the beam line. In the plane
transverse to the beam line the azimuthal angle $\phi$ defines the
directions of $\vec{\Pt}^{Jet}$ and $\vec{\Pt}^{\gamma}$. \\

\subsection{Introduction of some new  measurable physical observables
and $\Pt$ balance equation.}

In $pp \to \gamma + Jet + X$ events, we are going to study,
the main physical object will be a high $\Pt$ jet, to be detected
in the $|\eta| < 5$ region, and a direct photon, registered by
ECAL up to $|\eta| < 2.61$. In these events there will be a set of
particles mainly caused by beam remnants, i.e. by spectator partons
fragmentation, that are flying in the direction of a non-instrumented forward
part ($|\eta| > 5$) of the detector. Let us denote the total transverse
momentum of these non-observable particles as \begin{equation} \sum\limits_{i
\in |\eta| > 5} \vec{\Pt}^i \equiv \vec{\Pt}^{\eta > 5}.  \label{eq:sel1}
\end{equation} 

Among the particles with $|\eta| < 5$ there could be also neutrinos.
Their total momentum  will be denoted as
\begin{equation}
\sum\limits_{i \in |\eta| < 5} \vec{\Pt}_{(\nu)}^i
 \equiv \vec{\Pt}_{(\nu)}.
\label{eq:sel2}
\end{equation}
\vspace{-2mm}

\noindent
The sum of transverse momenta of these two kinds of non-detectable particles
will be denoted as $\Pt^{miss}$:
\vspace{-1mm}
\begin{eqnarray}
 \vec{\Pt}^{miss} = \vec{\Pt}_{(\nu)} + \vec{\Pt}^{\eta>5}.
\end{eqnarray}

\vspace{-2mm}

A high energetic jet can also contain neutrinos that may carry some
part of the total jet energy needed to be estimated from
simulation. From the total jet transverse
momentum $\vec{Pt}^{Jet}$  we shall separate the part that can be
measured in the detector, i.e. in the ECAL+HCAL and
muon systems.  Let us denote this part as $\vec{\Pt}^{jet}$ (small ``j''!).
 So, we shall
present the total jet transverse momentum $\vec{\Pt}^{Jet}$ as a sum of three
parts:  

1. $\vec{\Pt}^{Jet}_{(\nu)}$, containing the
contribution of neutrinos that belong to the jet, i.e.,
a non-detectable part of jet $\Pt$:
\vspace{-2mm}
\begin{eqnarray}
 \vec{\Pt}^{Jet}_{(\nu)} = \sum\limits_{i \in Jet} \vec{\Pt}_{(\nu)}^i.
\end{eqnarray}
~\\[-20pt]

2. $\vec{\Pt}^{Jet}_{(\mu)}$, containing the
contribution of jet muons to $\vec{\Pt}^{Jet}$. These muons
can give a weak signal in the calorimeters
but their energy can be measured in the muon system (in the
region of $|\eta|<2.4$ in the case of CMS geometry):
\vspace{-2mm}
\begin{eqnarray}
 \vec{\Pt}^{Jet}_{(\mu)} = \sum\limits_{i \in Jet} \vec{\Pt}_{(\mu)}^i.
\end{eqnarray}
~\\[-20pt]

Let us mention that due to the absence of the muon system and tracker
beyond the $|\eta|<2.4$ region,
there exists a part of $\Pt^{Jet}$ caused by muons with $|\eta|>2.4$
denoted as $\Pt^{Jet}_{(\mu,|\eta|>2.4)}$. This part
can be considered in some sense
as an analog of $\Pt^{Jet}_{(\nu)}$ since the only trace of its presence
would be weak MIP signals in ECAL and HCAL.

As for both points 1 and 2, let us say in advance that
the estimation of the average values of the neutrino and muon
contributions to $\Pt^{Jet}$ (see Section 4
and also Tables 1--8 of Appendix) has shown
that they are quite
negligible: about $0.35\%$ of $\la\Pt^{Jet}\ra_{all}$ is due to neutrinos and about
$0.25\%$ of $\la\Pt^{Jet}\ra_{all}$ --- to muons,
where $all$ means averaging over
all events including those without neutrinos in jets.

3. And finally, by means of $\vec{\Pt}^{jet}$ we denote
the part of $\vec{\Pt}^{Jet}$ which
includes all detectable particles of the jet
\footnote{We shall consider the issue of small $\Pt$ charged particles
contribution
into the total jet $\Pt$ while discussing the results of the full GEANT
 simulation (with account of the magnetic field effect) in our forthcoming
papers.}
, i.e.:
\vspace{-1mm}
\begin{eqnarray}
 \vec{\Pt}^{jet} =  \vec{\Pt}^{Jet}_{(HCAL+ECAL)} +\vec{\Pt}^{Jet}_{(\mu)} ,
\quad |\eta^{\mu}|<2.4
\end{eqnarray}

Thus, we can write (for the general case of $\eta$ values)
\vspace{-1.5mm}
\begin{eqnarray}
 \vec{\Pt}^{Jet}
=\vec{\Pt}^{jet}
+\vec{\Pt}^{Jet}_{(\nu)}
+\vec{\Pt}^{Jet}_{(\mu,|\eta^{\mu}|>2.4)}
\end{eqnarray}

In the case of $pp \to \gamma + Jet + X$ events
the particles detected in the $|\eta|<5$ region can originate
from the fundamental subprocesses (1a)
and (1b), that may be caused by LO diagrams, shown in Fig.~2 as well as
by NLO diagrams (like those in Fig.~3 that include ISR)
and also from the "underlying" event, of course.

As it was already mentioned in Section 2,
the final states of the fundamental subprocesses (1a) and (1b) may contain
additional
jets due to the ISR and final state radiation (FSR) caused by the higher
 order QCD
corrections to Feynman diagrams given in Fig.~1. To realize the calibration
idea (see Section 2.1), we need ''in situ'' selection of events with a good
balance of $\vec{\Pt}^{\gamma}$ and  the $\vec{\Pt}^{jet}$ part, measurable in
the detector. It means that to make a reasonable simulation, we need to
have a selected set of events with a small $\Pt_{(\nu)}$ (and, thus,
small $\Pt^{miss}$) as a model and we also have to find a way to select events
without additional
jets or with jets suppressed to the level of very small $\Pt$ mini-jets
or clusters.

For any event let us separate the particles in the $|\eta| < 5$ region into a
''$\gamma +Jet$'' system (here ''$Jet$'' denotes the jet with the highest
 $\Pt \geq 30 ~GeV/c$) having the total transverse momentum $\vec{\Pt}
^{\gamma +Jet}$ (see (4)) that may be different from:  \\[-10pt]
\begin{eqnarray}
\vec{\Pt}^{\gamma +jet} =
\vec{\Pt}^{\gamma} + \vec{\Pt}^{jet},
\end{eqnarray}
~\\[-10pt]
\noindent
in the case of neutrino presence in a jet,
 and a system of all other ($O$) particles  in the $|\eta| < 5$ region beyond
the ''$\gamma +Jet$'' system.
The total transverse momentum of this
system will be denoted as $\Pt^{O}$ and it is a sum of
$\Pt$  of additional mini-jets (or clusters) as well as of $\Pt$ of
single hadrons, photons and leptons in the $|\eta| < 5$ region. Since
neutrinos are present among these leptons, then
the difference of $\vec{\Pt}_{(\nu)}$ and $\vec{\Pt}^{Jet}_{(\nu)}$
gives us the value of the transverse momentum \\[-10pt]
\begin{eqnarray}
 \vec{\Pt}^{O}_{(\nu)} = \vec{\Pt}_{(\nu)} - \vec{\Pt}^{Jet}_{(\nu)}, \;\;
|\eta^{\nu}|<5
\end{eqnarray}

\noindent
carried out by neutrinos which do not belong to the jet but are
contained in the $|\eta| < 5$ region.


Let us denote a part of $\vec{\Pt}^O$, that can,
in principle, be measured in the detector, by $\vec{\Pt}^{out}$.
Thus, $\vec{\Pt}^{out}$ is a sum of $\Pt$ of other mini-jets or
clusters
(with $\Pt^{clust}$ smaller than $\Pt^{Jet}$) and of $\Pt$ of single
 hadrons ($h$), photons ($\gamma$) and electrons ($e$) with $|\eta| < 5$
 and muons ($\mu$) with $|\eta^\mu| < 2.4$.
Below for simplicity these mini-jets and clusters will be called just ''clusters''.
 So, $\vec{\Pt}^{out}$ is the following sum ($h,~\gamma,~e,~\mu \not\in$ Jet):
\begin{eqnarray}
 \vec{\Pt}^{out} =
\vec{\Pt}^{clust}
+\vec{\Pt}^{sing}_{(h)}
+\vec{\Pt}^{nondir}_{(\gamma)}
+\vec{\Pt}^{}_{(e)}+\vec{\Pt}^{O}_{(\mu, |\eta^\mu|<2.4)}; \quad  |\eta|<5
\end{eqnarray}

\noindent
And thus, finally, we have:
\begin{eqnarray}
 \vec{\Pt}^{O} =
\vec{\Pt}^{out}+\vec{\Pt}^{O}_{(\nu)}+\vec{\Pt}^{O}_{(\mu, |\eta^\mu|>2.4)}.
\end{eqnarray}

\noindent
With these notations the conservation law for the $\Pt$ component of the whole
''$\gamma + Jet$'' (where $\gamma$ is a direct photon) event is:
\begin{eqnarray}
\vec{\Pt}^{\gamma} +
\vec{\Pt}^{Jet} +
\vec{\Pt}^{O}+
\vec{\Pt}^{\eta>5} = 0
\end{eqnarray}

\noindent
with last three terms defined correspondingly by (11), (15) and (5).

\subsection{Definition of selection cuts for physical variables.}

\noindent
1. We select the events with one jet and one photon with\\[-5pt]
\begin{equation}
\Pt^{\gamma} \geq 40~ GeV/c~;\qquad \Pt^{Jet}\geq 30 \;GeV/c.
\label{eq:sc1}
\end{equation}
For most of our applications the jet is defined according to PYTHIA
jetfinding algorithm LUCELL.
The jet cone radius R in $\eta-\phi$ space
,counted from the jet initiator cell (ic), is
taken as $R_{ic}=((\Delta\eta)^2 + (\Delta\phi)^2)^{1/2}=0.7$
Below we shall also use the value of the jet radius, counted
from the center of gravity (gc) of the jet, i.e. $R_{gc}$.
The comparison
with UA1 jetfinding algorithm (taken from CMSJET
program of fast simulation [16]) is presented in [13, 14].

\noindent
2. To suppress the background processes, only the events with "isolated"
photons are taken. To do this, we restrict:

a) the value of the scalar sum of $\Pt$ of hadrons and other particles surrounding
a photon within a cone of $R^{\gamma}_{isol}=( (\Delta\eta)^2 + (\Delta\phi)^2)^{1/2}=0.7$
(``absolute isolation cut")\\[-7pt]
\begin{equation}
\sum\limits_{i \in R} \Pt^i \equiv \Pt^{isol} \leq \Pt_{CUT}^{isol};
\label{eq:sc2}
\end{equation}
\vspace{-2.6mm}

b) the value of a fraction (``relative isolation cut'')\\[-7pt]
\begin{equation}
\sum\limits_{i \in R} \Pt^i/\Pt^{\gamma} \equiv \epsilon^{\gamma} \leq 
\epsilon^{\gamma}_{CUT};
\label{eq:sc3}
\end{equation}

c) we accept only the events having no charged tracks (particles) 
with $\Pt>1~GeV/c$ within $R^{\gamma}_{isol}$ cone around the photon candidate.

\noindent
3. To be consistent with the application condition of the NLO
formulae, one should avoid an infrared dangerous region and take care of
$\Pt$ population in the region close to a photon (see [21-23]).
In accordance with [22] we also restrict the scalar sum of $\Pt$ of particles
 around a photon within a cone of a smaller radius $R_{singl} = 0.175 = 1/4
\,R_{isol}^{\gamma}$.

Due to this cut,
\begin{equation}
\sum\limits_{i \in R_{singl}} \Pt^i \equiv \Pt^{singl} \leq 2~ GeV/c,
~~~~~(i\neq ~\gamma-dir).
\label{eq:sc4}
\end{equation}
an ``isolated" photon with high $\Pt$ also becomes a ``single'' one within
an area of 8 towers (of 0.087x0.087 size according to CMS geometry)
 which surround the tower hitted by
it (analog of 3$\times$3 tower window algorithm).

\noindent
4. We also consider the structure of every event with the photon
candidate at a more precise level of 5x5 crystal cells window (size of one CMS
HCAL tower) with a cell size of 0.0175x0.0175. To suppress the background
events with photons resulting from high energetic $\pi^0-$, $\eta-$, $\omega-$
and $K_S^0-$ mesons,we apply in addition the following cut:

a) the ECAL signal can be considered as a candidate to be a direct photon if 
it fits inside the 3x3 ECAL crystal cell window with the highest $\Pt$ $\gamma/e$
in the center;

b) the value of a scalar sum of $\Pt$ ($\Pt^{sum}$) of
stable particles in the 5x5 crystal cell window in the region
out of a smaller 3x3 crystal cell window (typical size of photon shower
in ECAL found from GEANT simulation with CMSIM package), having the cell
with the direct photon candidate (the largest $\Pt$ $\gamma/e$)
as the central one,
should be restricted by $1 \, GeV/c$, i.e.
\begin{equation}
\Pt^{sum} \leq 1~ GeV/c;
\label{eq:sc6}
\end{equation}

c) we require the absence of a high $\Pt$ hadron
in this 5x5 crystal cell window (that means an imposing of an upper cut
on the HCAL signal at least in the one-tower area) around the direct photon:
\begin{equation} 
\Pt^{hadr} \leq 5~ GeV/c. 
\label{eq:sc5} 
\end{equation}

We can not reduce this value to, for example, 2-3 $GeV/c$, because
a hadron with $\Pt$ below 2-3 $GeV/c$ deposits most of its energy in ECAL and
may not reveal itself in HCAL.

\noindent   
5. We select the events with the vector $\vec{\Pt}^{Jet}$ being ``back-to-back" to
the vector $\vec{\Pt}^{\gamma}$ (in the plane transverse to the beam line) 
within $\dphi$ defined by equation:\\[-5pt]
\begin{equation}
\phigj=180^\circ \pm \Delta\phi \quad (\Delta\phi =15^\circ, 10^\circ, 5^\circ)
\label{eq:sc7}
\end{equation}
($5^\circ$ is a size of one CMS HCAL tower in $\phi$)
for the following definition of the angle $\phigj$  \\
\hspace*{2cm} $\vec{\Pt}^{\gamma}\vec{\Pt}^{Jet}=\Pt^{\gamma}\Pt^{Jet}\cdot cos(\phigj)$, ~~~
with ~$\Pt^{\gamma}=|\vec{\Pt}^{\gamma}|,~~\Pt^{Jet}=|\vec{\Pt}^{Jet}|$.  

\noindent
6. The initial state radiation  (ISR) manifests itself
as some final state cluster or mini-jet activity.
To suppress it, we impose a new cut condition that was not used earlier
in previous experiments: we choose the events that do not have any other
jet-like or cluster high $\Pt$ activity, i.e. $\Pt^{clust}$ (taking
the cluster cone $R_{clust}(\eta,\phi)=0.7$), being higher than some threshold
$\Pt^{clust}_{CUT}$ value, i.e. we select the events with\\[-10pt]
\begin{equation}
\Pt^{clust} \leq \Pt^{clust}_{CUT},
\label{eq:sc8}
\end{equation}
where clusters are found by one and the same jetfinder used to find the jet
in the event.

\noindent
7. We limit the value of modulus of the vector sum of $\vec{\Pt}$ of all
particles, except those in the \gpj system, that fit into the region covered by
ECAL and HCAL
(i.e. the cells ``beyond the jet and photon'' regions):
\begin{equation}
\left|\sum_{i\not\in Jet,\gamma-dir}\vec{\Pt}^i\right| \equiv \Pt^{out} \leq \Pt^{out}_{CUT}, ~~|\eta|<5
\label{eq:sc9}
\end{equation}

\noindent
The importance of $\Pt^{out}_{CUT}$ and $\Pt^{clust}_{CUT}$
parameters  to reduce the background
will be demonstrated in the forthcoming papers [13--15].

Below the selection cuts 1 -- 7 will be referred as
``Selection 1``. The last two of them, 6 and 7, are new criteria,
not used in previous experiments. In addition to them one more
new object, named an ''isolated jet'', will be introduced.

\noindent
8. To do this we also involve
a new requirement of ``jet isolation'',
i.e. the presence of a ``clean enough'' (in the sense of limited $\Pt$
activity) region inside the ring (of $\Delta R=0.3$ size) around the
jet.  Following this picture we restrict the value of the ratio of the scalar sum
of transverse momenta of particles belonging to this ring, i.e.\\[-5pt]
\begin{equation}
\Pt^{ring}/\Pt^{\gamma} \equiv \epsilon^{jet} \leq 2\%
, \quad {\rm where ~~~~ }
\Pt^{ring}=\sum\limits_{\footnotesize i \in 0.7<R<1} |\vec{\Pt}^i|.
\label{eq:sc10}
\end{equation}
~\\[-4pt]
The set of events that 
pass under the cuts 1 -- 8 will be called as ``Selection 2''.

\noindent
9. In  the following ``Selection 3'' we shall keep only those events in which
one and the same jet (i.e. up to good accuracy having the same values
of $\Pt^{Jet}, ~R^{jet}$ and $\Delta\phi$) is found simultaneously by
every of two jetfinders used here: UA1 and LUCELL.
For these jets (and also clusters) we require the following conditions:
\begin{eqnarray}
\Pt^{Jet}>30~GeV/c, \qquad \Pt^{clust}< \Pt^{clust}_{CUT},\qquad
\dphi<15^\circ (10^\circ,~5^\circ), \qquad \epsilon^{jet} \leq 2\%
\end{eqnarray}

The exact values of cut parameters, i.e. $\Pt^{isol}_{CUT}$,
$\epsilon^{\gamma}_{CUT}$, $\epsilon^{jet}$, $\Pt^{clust}_{CUT}$, $\Pt^{out}_{CUT}$,
 will be specified below, since they may be
different, for instance, for various $\Pt^{\gamma}$-intervals
(being more loose for higher  $\Pt^{\gamma}$).

\noindent
10. As we have already mentioned in Section 3.1, one can expect
reasonable results of the calibration procedure modeling only by using
a set of selected events with a small value of $\Pt^{miss}$. So, we also use
the following cut:
\begin{eqnarray}
\Pt^{miss}~\leq \Pt^{miss}_{CUT}.
\label{eq:sc11}
\end{eqnarray}
Due to this reason in Section 4, we shall study the influence of
$\Pt^{miss}$ parameter on the selection of events with a reduced value
of $\Pt^{Jet}_{(\nu)}$.
The aim of the event selection with a small value of $\Pt^{Jet}_{(\nu)}$
is quite obvious: we need a set of events with a reduced value of
$\Pt^{Jet}$ uncertainty due to possible presence of a non-detectable
neutrino contribution to a jet.

To conclude this section, let us rewrite
the basic $\Pt$-balance equation (16) of the previous section
by means of notations introduced here in the form
more suitable to present results in further papers [12--15]. For this
purpose we shall rewrite equation (16) in the following scalar form: \\[-20pt]
\begin{eqnarray}
\frac{\Pt^{\gamma}-\Pt^{Jet}}{\Pt^{\gamma}}=(1-cos\dphi) 
+ \Db/\Pt^{\gamma}, \label{eq:sc12}
\end{eqnarray} where
$\Db\equiv (\vec{\Pt}^{O}+\vec{\Pt}^{|\eta|>5)})\cdot \vec{n}^{Jet}$ ~~~~ with ~~
$\vec{n}^{Jet}=\vec{\Pt}^{Jet}/\Pt^{Jet}$.

As it will be shown in [12--14], the first term in the right-hand part of
the equation (29) is negligibly small and tends to decrease more with
a growth of the energy. So, the main source of the $\Pt$ disbalance
in the \gpj system is a term $\Db/\Pt^{\gamma}$.

\section{ESTIMATION OF NON-DETECTABLE PART OF $\Pt^{Jet}$}       

In Section 3.1 we have separated the transverse momentum of the jet,
i.e. $\Pt^{Jet}$, into two parts:
a detectable one $\Pt^{jet}$
 and a non-measurable part, consisting of $\Pt^{Jet}_{(\nu)}$
(see (8)) and $\Pt^{Jet}_{(\mu,|\eta|>2.4)}$ (see (11))
\footnote{Firstly we shall consider the case of switched off decays of
$\pi^{\pm}$ and $\;K^{\pm}$ mesons (according to the PYTHIA default
agreement, $\pi^{\pm}$ and $\;K^{\pm}$ mesons are stable).}.
In the same way we have done analogous separation according to equation (15)
of the transverse momentum of other particles, i.e. $\Pt^O$,
excluding direct photon (or candidate to be
detected as a direct photon), into detectable part $\Pt^{out}$
and non-measurable part consisting of $\Pt^O_{(\nu)}$
(see (13)) and $\Pt^O_{(\mu,|\eta|>2.4)}$ (see (15)).

Here we present an estimation of averaged values of transverse momenta
of the total $\Pt^{Jet}$ carried out by non-detectable particles.
For this aim we use a bank of the signal \gpj events generated for
three intervals of $\Ptg$ with the restrictions (17) -- (24) and the following
cuts are fixed as follows:
\begin{equation}
 \Pt_{CUT}^{isol}=20 ~GeV/c,~~\epsilon^{\gamma}_{CUT}=15\%,
~~\dphi=15^\circ,~~\Pt^{clust}_{CUT}=30~ GeV/c.
\end{equation}
No restriction for the $\Pt^{out}$ value was done.
The results of analysis of these events are presented in Figs.~4 and 5.

The first row of Fig.~\ref{fig20-22}
contains  $\Pt^{miss}$ spectra in the \gpj events for different
$\Ptg$ intervals. Their practical independence
(up to the good accuracy) on $\Ptg$ is clearly seen.

In the second row of Fig.~\ref{fig20-22} we present the spectra of
$\Pt^{miss}$ for the events (denoted as $\Pt^{Jet}_{(\nu)}>0$)
having a non-zero $\Pt^{Jet}_{(\nu)}$
component in $\Pt^{Jet}$. For these figures the $\Pt^{miss}$ spectrum
dependence on the direct photon $\Pt^{\gamma}$ (that is equal,
approximately, to $\Pt^{Jet}$) is seen.  So, the spectra tails as well as
the mean values
are shifting to a large $\Pt^{miss}$ region with $\Pt^{Jet}$ growth.
 At the same time a peak position
remains in the region of $\Pt^{miss}<5 ~GeV/c$. From the comparison of the
number of entries in the second row plots of Fig.~\ref{fig20-22} with
those in the first row it can be concluded that the part of events with the jet
having the non-zero neutrinos contribution ($\Pt^{Jet}_{(\nu)}>0$) has the same
size of about $3.3\%$ in all $\Ptg$ intervals.

The same spectra of $\Pt^{miss}$ for events with $\Pt^{Jet}_{(\nu)}>0$
show what amount of these events would remain after imposing a cut on
$\Pt^{miss}$ in every $\Ptg$ interval.
The important thing here is that the reduction of the number of events
with $\Pt^{Jet}_{(\nu)}>0$~ in every $\Ptg$ interval leads to reducing
the mean value of  $\Pt^{Jet}_{(\nu)}$, i.e. a value averaged over all
collected events.
This value, found from PYTHIA generation, serves as a model correction
$\Delta_\nu$ and it has to be estimated for the proper determination
of the total $\Pt^{Jet}$ from the measurable part $\Pt^{jet}$: $\Pt^{Jet}=
\Pt^{jet} + \Delta_\nu$, where $\Delta_\nu=\la\Pt^{Jet}_{(\nu)}\ra_{all\; events}$.

The effect of general $\Pt^{miss}_{CUT}$ imposing in each event of our
sample is shown in the third row of Fig.~\ref{fig20-22}. The upper cut
$\Pt^{miss}_{CUT}=1000 ~GeV/c$, as it is seen from the second row
pictures, means the absence of any upper limit on $\Pt^{Jet}_{(\nu)}$.
The most important information that the value of the neutrinos $\Pt$ inside the jet,
being averaged over all events, can reach the value of 
$\Pt^{Jet}_{(\nu)}\approx 1~GeV/c$
at $\Ptg\geq 300~ GeV/c$ comes from the right-hand plot of the third row in Fig.~4.
From the comparison of the plots from the second row with the corresponding plots
\footnote{That includes the values of $\Pt^{miss}_{CUT}$ and the corresponding
number of entries remained after $\Pt^{miss}_{CUT}$ imposing as well as the mean
value of $\Pt^{Jet}_{(\nu)}$, denoted as ``Mean'' (being equal to an averaged
$\la \Pt^{Jet}_{(\nu)}\ra$ value over the number of the remained entries.}
from the third row
we see that the first cut $\Pt^{miss}_{CUT}=20 ~GeV/c$ for the first
$40<\Ptg<50 ~GeV/c$ interval reduces the number of entries by less than
$0.5\%$  and the mean value of $\Pt^{Jet}_{(\nu)}$ --- by less than $10\%$.
A more restrictive cut $\Pt^{miss}_{CUT}=5 ~GeV/c$ reduces the value of
 $\la \Pt^{Jet}_{(\nu)}\ra$ by three times and leads to approximate twofold
drop of the number of events.

From these Figures we see that for the interval $300<\Ptg<360$ the number of
events with jets including neutrino (second row)
is about $3.3\%$ (Entries=3001) of the total number
of the generated \gpj events (Entries=89986). A very restrictive
$\Pt^{miss}_{CUT}$=5 $GeV/c$ cut leads to the reduction
factor for $\la\Pt^{Jet}_{(\nu)}\ra$ of about 50. As it is seen from the
plot in the bottom right-hand corner
of Fig.~\ref{fig20-22}, the $\Pt^{Jet}_{(\nu)}$ spectrum for the remaining
 events (Entries=57475) finishes at
 $\Pt^{Jet}_{(\nu)}=10~ GeV/c$ and sharply peaks at
$\Pt^{Jet}_{(\nu)}=0$. The averaged value of $\Pt^{Jet}_{(\nu)}$ under this
 peak is equal to 0.022 $GeV/c$. So, with this cut on $\Pt^{miss}$ the neutrinos
give a negligible contribution to $\Pt^{Jet}$.

At the same time we see that application of the moderate cut
$\Pt^{miss}_{CUT}=10 ~GeV/c$ for
$300<\Pt^{Jet}<360~GeV/c$ interval strongly reduces
(by 20 times) the mean value of $\Pt^{Jet}_{(\nu)}$ (from $1~ GeV/c$ down
to $<\Pt^{Jet}_{(\nu)}>=0.05~GeV/c$) at about $10\%$ reduction of
the total number of events in this $\Pt^{Jet}$ (or $\Ptg$) interval.

In the case of $100<\Pt^{Jet}<120~GeV/c$ interval,
as we see from the third row of Fig.~\ref{fig20-22},
the same  cut $\Pt^{miss}_{CUT}=10 ~GeV/c$ reduces
the mean value of $\Pt^{Jet}_{(\nu)}$ by 5 times (from $0.5~ GeV/c$ down
to $<\Pt^{Jet}_{(\nu)}>=0.09~GeV/c$) with the same $10\%$ reduction of the
total number of events.

It should be noted that in the less dangerous (from the point of view of the size of
\begin{center}
\begin{figure}[htbp]
\vspace{-2.9cm}
\hspace{.0cm} \includegraphics[width=13.5cm]{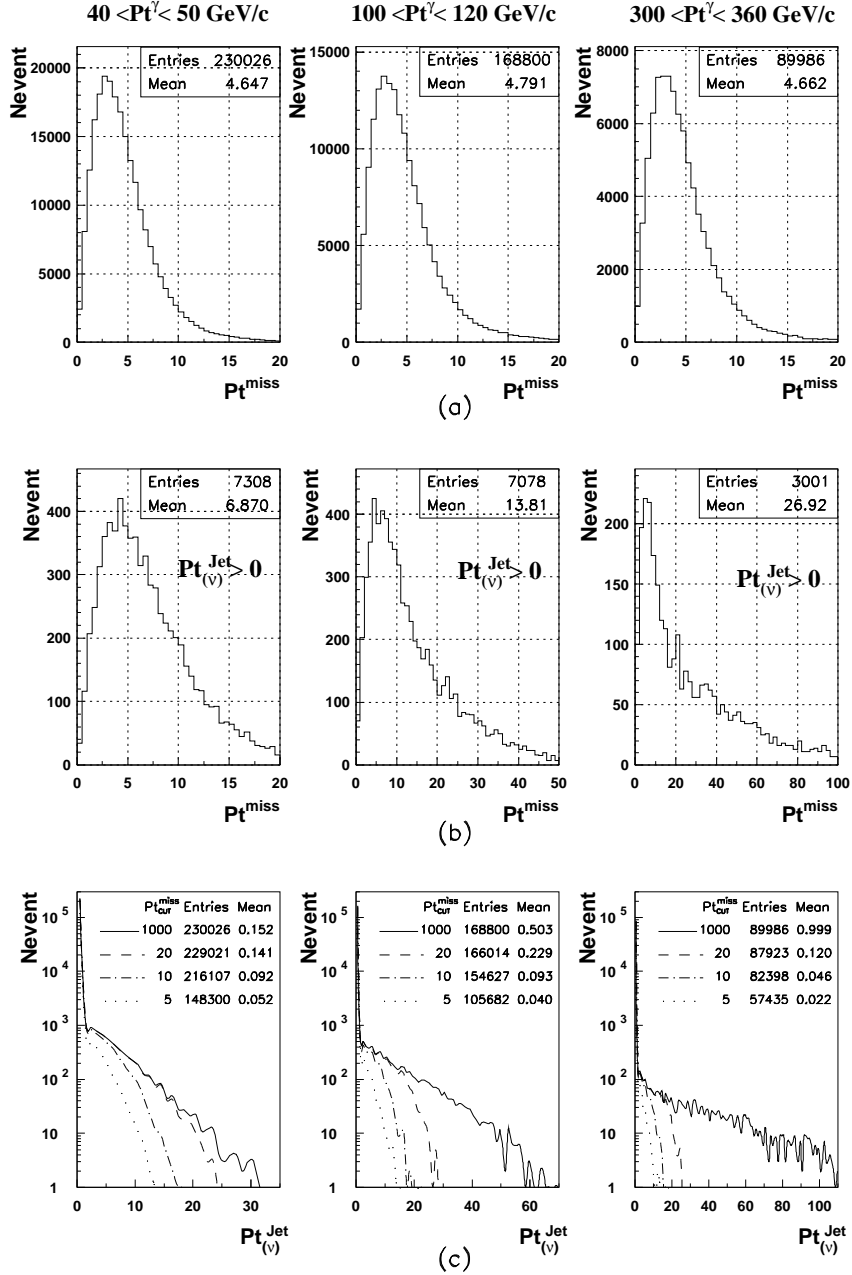}
\vspace{-0.5cm}
\caption{\hspace*{0.0cm} a) $\Pt^{miss}$ spectra in all events;
b) $\Pt^{miss}$ spectra in events having jets with non-zero $\Pt$
neutrinos, i.e. $\Pt^{Jet}_{(\nu)}>0$;~ c) $\Pt^{Jet}_{(\nu)}$ spectra
behavior
for different values of $\Pt^{miss}_{CUT}$ values in various
$\Pt^{Jet}(\approx\Pt^\gamma)$ intervals.}
\label{fig20-22}
\end{figure} 
\begin{figure}[htbp]
\vspace{-2.9cm}
\hspace{.0cm} \includegraphics[width=13.5cm]{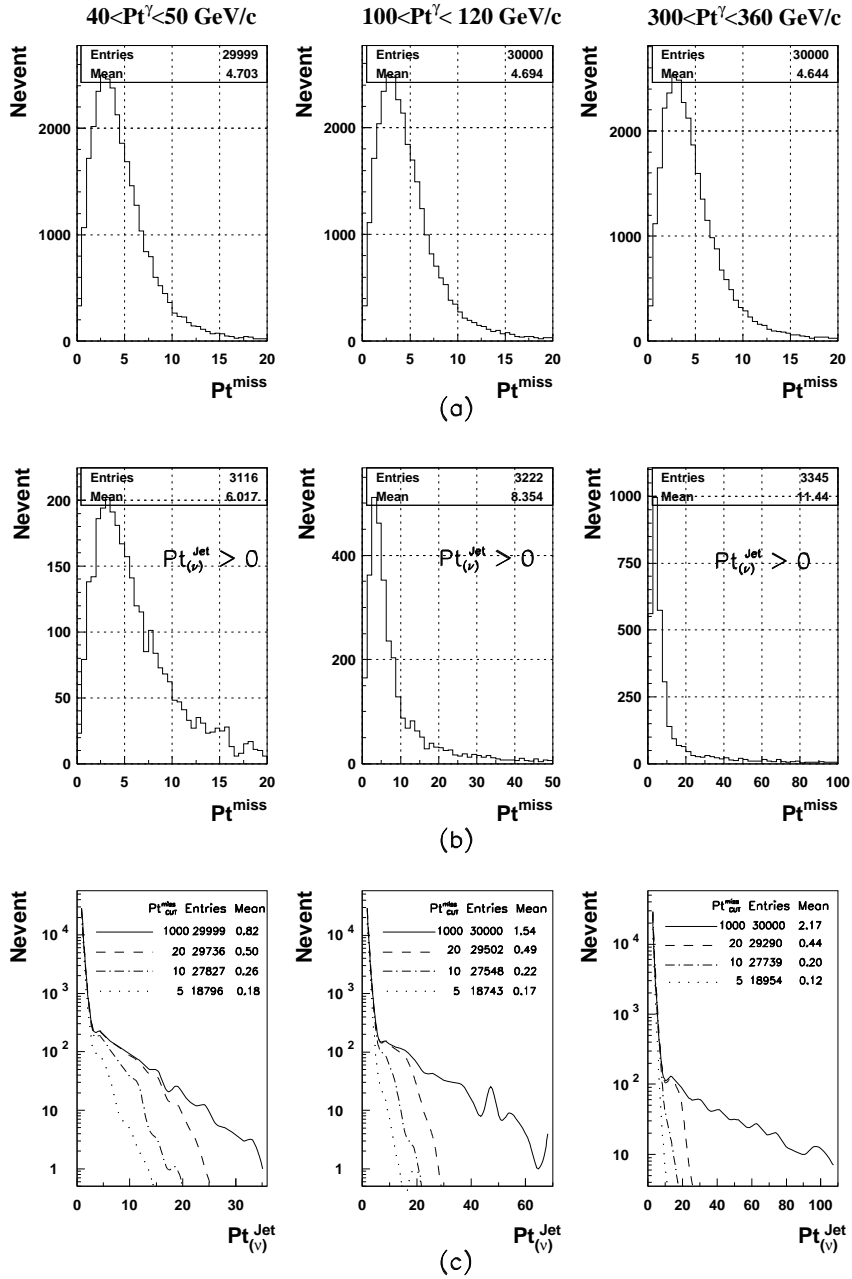}
\vspace{-0.5cm}
\caption{\hspace*{0.0cm} a) $\Pt^{miss}$ spectra in all events;
b) $\Pt^{miss}$ spectra in events having jets with non-zero $\Pt$
neutrinos, i.e. $\Pt^{Jet}_{(\nu)}>0$;~ c) $\Pt^{Jet}_{(\nu)}$ spectra
 behavior
for different values of $\Pt^{miss}_{CUT}$ values in various
$\Pt^{Jet}(\approx\Pt^\gamma)$ intervals.
$K^{\pm}-$decays are allowed inside the solenoid of $R=129~cm$ and $L=317~cm$.}
\label{fig23}
\end{figure} 
\end{center}

\noindent
neutrino $\Pt$ content in a jet) $40<\Pt^{Jet}<50~GeV/c$ interval we have already
 a very small mean value of $\Pt^{Jet}_{(\nu)}$ equal to $0.152 ~GeV/c$
 without imposing any $\Pt^{miss}_{CUT}$.


The analogous (to neutrino) situation holds for $\Pt^{Jet}_{(\mu)}$
contribution.

The detailed information about the values of non-detectable
$\Pt^{Jet}_{(\nu)}$, averaged over all events
(no cut on $\Pt^{miss}$ was used), as well as about mean values of
muons $\Pt$ from jet
$\Pt^{Jet}_{(\mu)}$, is presented
in Tables 1--8 of Appendix for the sample of events with jets which are
completely contained in the barrel region of HCAL ($|\eta^{jet}|<1.4$, ``HB-events'',
see [12]).
In these tables the ratio of the number of events with non-zero
$\Pt^{Jet}_{(\nu)}$
to the total number of events is denoted by $R^{\nu \in Jet}_{event}$ and
the ratio of the number of events with non-zero $\Pt^{Jet}_{(\mu)}$
to the total number of events is denoted by $R^{\mu \in Jet}_{event}$.
The $\Pt^{miss}$ quantity
in events with $\Pt^{Jet}_{(\nu)}>0$ is denoted in these tables as
$\Pt^{miss}_{\nu \in Jet}$ and is given there for four $\Ptg$ intervals
and other $\Pt^{clust}_{CUT}$ values than in the second row plots
of Figs.~4 and 5. From Tables 1--8 we see that the averaged of
$\Pt^{miss}$ value being calculated by using of only the
events with $\Pt^{Jet}_{(\nu)}>0$, i.e.
$\la\Pt^{miss}_{\nu \in Jet}\ra$, is about 7 $GeV/c$ for $40<\Ptg<50 ~GeV/c$
interval. It increases to about 32 $GeV/c$ for $300<\Ptg<360 ~GeV/c$.
It should be noted that the averaged values of the modulus of
$\Pt^{Jet}_{(\nu)}$ (see
formula (8)) presented in the second and third lines of Tables 1--6
from the Appendix, coincide with the averaged values of
$\Delta_\nu$ up to the three digits due to practical collinearity of
$\vec{\Pt}^{Jet}$ and $\vec{\Pt}^{jet}$ vectors, i.e. $<\!\!\Pt^{Jet}_{(\nu)}\!\!>=
<\!\!\Delta_\nu\!\!>$.

Tables 1--8 contain the lines with an additional information on
the numbers of the \gpj events with the jets produced by $c-$ and $b-$ quarks, i.e.
$Nevent_{(c)}$ and $Nevent_{(b)}$, given for the integrated luminosity 
$L_{int}=3~ fb^{-1}$, respectively.
There are also lines that show a ratio (``$29sub/all$'')
of the number of the events caused by gluonic (``Compton-like'')
subprocess (1a) to the number of events due to the sum of
(1a) and (1b) subprocesses.

Below follows the line containing the averaged values of the jet radius 
$<\!\!R_{jet}\!\!>$.

The value of the difference of the jet measurable
part transverse momentum
$\Pt^{jet}$  and of the total jet $\Pt^{Jet}$,
averaged over all events,
 i.e. $\la\Pt^{Jet}-\Pt^{jet}\ra$,
is presented in second lines of Tables 1--8
of Appendix in $GeV/c$ units. This value has a sense of
correction $\Delta_{\nu}$ that should be applied to $\Pt^{jet}$
in order to take into account $\Pt$ carried off by non-detectable particles.

It was already mentioned in Introduction that we are planning to carry out a more
detailed analysis basing on GEANT package and taking into account weak decays
 of $\pi^\pm$ and $K^\pm$ mesons.
To have an idea what changes can be expected, we shall consider 
only the case of allowed $K^{\pm}$ decays (as
the main source of neutrinos and muons).
The averaged values of $\Pt^{Jet}_{(\nu)}$ for different $\Ptg$-intervals
with switched on $K^{\pm}$ decays
are given in Fig.~\ref{fig23} with the same
meaning of all notations. Here $K^{\pm}$ decays are allowed inside
the solenoid volume with the barrel radius $R_B=129~ cm$ and the distance
from the interaction vertex to End-cap along Z-axis $L=317~cm$
(CMS geometry).

From this Figure we see that in the case of allowed $K^{\pm}$ decays,
the $\Pt^{miss}$ spectrum for all events (compare the first rows in Fig.~4
and Fig.~5) practically does not change with $\Pt^{Jet}(\approx \Ptg)$
as well as the mean value of $\Pt^{miss}$.
At the same time the $\Pt^{miss}$ spectra change for events that
contain neutrinos in the jet (second row of Fig.~5) quite noticeably.
It should be noted that the amount of such
events grows up to $10\%$ as compared with $3\%$ in 
the case considered in Fig.~4,
but the mean values of
$\Pt^{miss}$ do not grow so strongly with $\Pt^{\gamma}$ as it was seen in Fig.~4.
The mean value of $\Pt^{miss}$ changes only twice
from $6.0~ GeV/c$ for the 
interval $40\!<\!\Ptg\!<\!50$ to $11.4~ GeV/c$
for the interval $300<\Ptg<360$.
Now we compare the third row pictures in Figs.~4 and 5.
We see that the mean value
of $\Pt^{Jet}_{(\nu)}$, carried out by
neutrinos of the  jet grows up from
the value of about $\la\Pt^{Jet}_{(\nu)}\ra\approx0.8~GeV/c$ at
$40\!<\!\Ptg\!<\!50~ GeV/c$ to the value of
$\la\Pt^{Jet}_{(\nu)}\ra\approx2.2~GeV/c$ at
$300\!<\!\Ptg\!<\!360~ GeV/c$, i.e. it changes as $2\%\to0.7\%$
of $\Pt^{Jet}$.
From the same pictures of Fig.~5 we see that the general cut
$\Pt^{miss}_{CUT}=20 ~GeV/c$ would reduce the contribution of neutrinos into
$\Pt^{Jet}$ to the value $\la\Pt^{Jet}_{(\nu)}\ra\approx0.5~GeV/c$ in all
$\Pt^{Jet}$ 
intervals, while the cut $\Pt^{miss}_{CUT}=10 ~GeV/c$ would lead to
the value of $\la\Pt^{Jet}_{(\nu)}\ra\approx0.20-0.26 ~GeV/c$.
(that is quite acceptable)  with $\approx 9\%$ reduction of the event number.

\section{SUMMARY}                                                   

The possibility of the jet energy scale setting and hadron calorimeter
calibration at LHC energies by using the \gpj process is considered for the
case of low luminosity ($10^{33}~cm^2 s^{-1}$).

The initial state radiation (ISR) as the main source of the $\Pt$
 disbalance in the \gpj system
is discussed.
New variables that enter the $\Pt$-balance equation are considered.
The new cuts (see Section 3) $\Pt^{out}_{CUT}$~ and ~ $\Pt^{clust}_{CUT}$
(in addition to the cut on $\phigj$ angle used previously in other
experiments) as well as a new object of ``isolated jet'' are introduced
here. The consequences of their variation and the choice
of their most preferable values to select the events with a good
\ptgj balance will be discussed in [12--15].


The values of the non-detectable part of $\Pt^{Jet}$ caused by neutrinos
are estimated for different $\Pt^{Jet}$ intervals. It is found that
$\Pt^{\gamma}$ and $\Pt^{Jet}$ balance can be influenced by
neutrino energy leakage from the jet.
The $\Pt^{miss}_{CUT}=10 ~GeV/c$ is proved to be sufficient for reducing the
jet energy leakage caused by neutrinos to an
acceptable level with about of $9\%~$ loss of events.

After the detailed study of neutrinos and muons contribution to $\Pt^{Jet}$ done here,
the following papers will be concentrated on the contribution of
hadrons, photons and electrons.

We would like to emphasize once more that  the values
of selection cuts, given here, are not a dogma for us. Our aim is to present
an estimation of a number of events that can be selected in some unit of
time chosen here as one month of LHC continuous operation
(i.e.  $3000\,pb^{-1}=3\, fb^{-1}$). In future calibration "in situ"
one can collect these \gpj events and in parallel 
classify them according to different Selection criteria (1, 2 and 3; see Section 3.2)
 for increasing the degree of accuracy.

\section{ ACKNOWLEDGMENTS}                                              

We are greatly thankful to D.~Denegri for having offered this theme to study,
fruitful discussions and permanent support and encouragement.
It is a pleasure for us
to express our recognition for helpful discussions to P.~Aurenche,
M.~Dittmar, M.~Fontannaz, J.Ph.~Guillet, M.L.~Mangano, E.~Pilon,
H.~Rohringer, S.~Tapprogge and especially to J.~Womersley for supplying us with
the preliminary version of paper [1].


\newpage
\begin{table}[h]
{\bf APPENDIX}\\[10pt]                                             
\begin{center}
\vskip-0.5cm
\large{ $40 < \Pt^{\gamma} < 50 ~GeV/c$}\\[20pt]
\normalsize
\caption{Selection 1. $\phigj=180^\circ \pm 15^\circ$. UA1 algorithm. ~$L_{int}=3 ~fb^{-1}$}
\vskip0.2cm
\begin{tabular}{||c||c|c|c|c|c|c|c|c|c|c||}
\hline
\hline
$Pt^{clust}_{CUT}$ &$\quad$  30 $\quad$&$\quad$  20 $\quad$&$\quad$ 15 $\quad$&$\quad$  10 $\quad$&$\quad\ $   5 $\quad\ $\\\hline
\hline
$Pt^{jet}$                &   43.021&   42.771&   42.679&   42.755&   43.202\\\hline
$\nuj$                    &    0.168&    0.167&    0.161&    0.160&    0.127\\\hline
$Pt_{(\nu)}^{Jet}$        &    0.169&    0.168&    0.162&    0.161&    0.128\\\hline
$R_{event}^{\nu\in Jet}$  &    0.033&    0.033&    0.033&    0.033&    0.027\\\hline
$Pt_{(\mu)}^{Jet}$        &    0.100&    0.099&    0.096&    0.099&    0.087\\\hline
$R_{event}^{\mu\in Jet}$  &    0.018&    0.018&    0.017&    0.017&    0.014\\\hline
$Pt^{miss}$               &    4.551&    4.511&    4.470&    4.399&    4.134\\\hline
$Pt^{miss}_{\nu\in Jet}$  &    7.054&    6.942&    6.843&    6.777&    6.576\\\hline
Nevent$_{(c)}$            &   312191&   287694&   253628&   180811&    40334\\\hline
Nevent$_{(b)}$            &    40098&    36223&    30495&    20689&     3639\\\hline
$29sub/all$               &     0.92&     0.91&     0.91&     0.91&     0.90\\\hline
$R_{jet}$                 &     0.60&     0.60&     0.60&     0.60&     0.59\\\hline
                   Entries&    56532&    52588&    46991&    34426&     8421\\\hline
\hline
\end{tabular}
\vskip0.3cm
\caption{Selection 1. $\phigj=180^\circ \pm 15^\circ$. LUCELL algorithm. ~$L_{int}=3 ~fb^{-1}$ }
\vskip0.2cm
\begin{tabular}{||c||c|c|c|c|c|c|c|c|c|c||}
\hline\hline
$Pt^{clust}_{CUT}$ &$\quad$  30 $\quad$&$\quad$  20 $\quad$&$\quad$ 15 $\quad$&$\quad$  10 $\quad$&$\quad\ $   5 $\quad\ $\\\hline
\hline
$Pt^{jet}$                &   43.253&   43.000&   42.949&   43.026&   43.408\\\hline
$\nuj$                    &    0.168&    0.165&    0.160&    0.156&    0.121\\\hline
$Pt_{(\nu)}^{Jet}$        &    0.169&    0.166&    0.161&    0.157&    0.121\\\hline
$R_{event}^{\nu\in Jet}$  &    0.033&    0.033&    0.033&    0.032&    0.027\\\hline
$Pt_{(\mu)}^{Jet}$        &    0.103&    0.100&    0.098&    0.093&    0.094\\\hline
$R_{event}^{\mu\in Jet}$  &    0.018&    0.018&    0.017&    0.017&    0.015\\\hline
$Pt^{miss}$               &    4.556&    4.510&    4.474&    4.382&    4.104\\\hline
$Pt^{miss}_{\nu\in Jet}$  &    7.027&    6.915&    6.834&    6.745&    6.595\\\hline
Nevent$_{(c)}$            &   304172&   277451&   241228&   164132&    36021\\\hline
Nevent$_{(b)}$            &    39256&    34740&    29248&    18937&     3167\\\hline
$29sub/all$               &     0.92&     0.91&     0.91&     0.90&     0.90\\\hline
$R_{jet}$                 &     0.65&     0.65&     0.65&     0.64&     0.63\\\hline
                   Entries&    54922&    50723&    44738&    31455&     7751\\\hline
\hline
\end{tabular}
\end{center}
\end{table}

\begin{table}
\begin{center}
\large{ $100 < \Pt^{\gamma} < 120 ~GeV/c$}\\[20pt]
\normalsize
\caption{Selection 1. $\phigj=180^\circ \pm 15^\circ$. UA1 algorithm.  ~$L_{int}=3 ~fb^{-1}$}
\vskip0.2cm
\begin{tabular}{||c||c|c|c|c|c|c|c|c|c|c||}
\hline
\hline
$Pt^{clust}_{CUT}$ &$\quad$  30 $\quad$&$\quad$  20 $\quad$&$\quad$ 15 $\quad$&$\quad$  10 $\quad$&$\quad\ $   5 $\quad\ $\\\hline
\hline
$Pt^{jet}$                &  102.627&  104.675&  105.575&  106.329& 106.917\\\hline
$\nuj$                    &    0.546&    0.538&    0.523&    0.501& 0.488\\\hline
$Pt_{(\nu)}^{Jet}$        &    0.548&    0.539&    0.525&    0.502& 0.489\\\hline
$R_{event}^{\nu\in Jet}$  &    0.044&    0.042&    0.040&    0.038& 0.034\\\hline
$Pt_{(\mu)}^{Jet}$        &    0.258&    0.249&    0.234&    0.228& 0.216\\\hline
$R_{event}^{\mu\in Jet}$  &    0.023&    0.022&    0.021&    0.019& 0.019\\\hline
$Pt^{miss}$               &    5.166&    5.139&    5.102&    5.053& 4.913\\\hline
$Pt^{miss}_{\nu\in Jet}$  &   14.245&   14.512&   14.817&   14.831& 15.412\\\hline
Nevent$_{(c)}$            &    18289&    13417&    10124&     6149& 1051\\\hline
Nevent$_{(b)}$            &     2887&     1989&     1435&      779& 113\\\hline
$29sub/all$               &     0.89&     0.89&     0.88&     0.88& 0.86\\\hline
$R_{jet}$                 &     0.61&     0.61&     0.61&     0.61& 0.60\\\hline
                   Entries&    63316&    48178&    37512&    23472& 4467\\\hline
\hline
\end{tabular}
\vskip0.3cm
\caption{Selection 1. $\phigj=180^\circ \pm 15^\circ$. LUCELL algorithm. ~$L_{int}=3 ~fb^{-1}$ }
\vskip0.2cm
\begin{tabular}{||c||c|c|c|c|c|c|c|c|c|c||}
\hline
\hline
$Pt^{clust}_{CUT}$ &$\quad$  30 $\quad$&$\quad$  20 $\quad$&$\quad$ 15 $\quad$&$\quad$  10 $\quad$&$\quad\ $   5 $\quad\ $\\\hline
\hline
$Pt^{jet}$                &  103.378&  105.266&  106.137&  106.938&  107.216\\\hline
$\nuj$                    &    0.549&    0.544&    0.524&    0.475&    0.491\\\hline
$Pt_{(\nu)}^{Jet}$        &    0.552&    0.546&    0.525&    0.477&    0.492\\\hline
$R_{event}^{\nu\in Jet}$  &    0.045&    0.043&    0.041&    0.037&    0.034\\\hline
$Pt_{(\mu)}^{Jet}$        &    0.260&    0.249&    0.240&    0.223&    0.198\\\hline
$R_{event}^{\mu\in Jet}$  &    0.023&    0.022&    0.021&    0.019&    0.017\\\hline
$Pt^{miss}$               &    5.168&    5.136&    5.110&    5.010&    4.897\\\hline
$Pt^{miss}_{\nu\in Jet}$  &   14.169&   14.506&   14.527&   14.442&   15.832\\\hline
Nevent$_{(c)}$            &    17309&    12498&     9257&     5137&      984\\\hline
Nevent$_{(b)}$            &     2704&     1866&     1308&      620&      102\\\hline
$29sub/all$               &     0.89&     0.89&     0.88&     0.87&     0.85\\\hline
$R_{jet}$                 &     0.65&     0.65&     0.65&     0.65&     0.64\\\hline
                   Entries&    59683&    44691&    34139&    20072&     4019\\\hline
\hline
\end{tabular}
\end{center}
\end{table}

\begin{table}
\begin{center}
\large{ $200 < \Pt^{\gamma} < 240 ~GeV/c$}\\[20pt]
\normalsize
\caption{Selection 1. $\phigj=180^\circ \pm 15^\circ$. UA1 algorithm.  ~$L_{int}=3 ~fb^{-1}$}
\vskip0.2cm
\begin{tabular}{||c||c|c|c|c|c|c|c|c|c|c||}
\hline
\hline
$Pt^{clust}_{CUT}$ &$\quad$  30 $\quad$&$\quad$  20 $\quad$&$\quad$ 15 $\quad$&$\quad$  10 $\quad$&$\quad\ $   5 $\quad\ $\\\hline
\hline
$Pt^{jet}$                &  211.973&  213.370&  214.124&  214.874&  215.511\\\hline
$\nuj$                    &    0.886&    0.874&    0.823&    0.768&    0.639\\\hline
$Pt_{(\nu)}^{Jet}$        &    0.889&    0.877&    0.825&    0.770&    0.640\\\hline
$R_{event}^{\nu\in Jet}$  &    0.039&    0.038&    0.037&    0.035&    0.033\\\hline
$Pt_{(\mu)}^{Jet}$        &    0.420&    0.399&    0.360&    0.338&    0.344\\\hline
$R_{event}^{\mu\in Jet}$  &    0.020&    0.019&    0.019&    0.017&    0.016\\\hline
$Pt^{miss}$               &    5.558&    5.529&    5.444&    5.368&    5.130\\\hline
$Pt^{miss}_{\nu\in Jet}$  &   24.382&   24.517&   23.705&   23.182&   20.955\\\hline
Nevent$_{(c)}$            &     1081&      753&      547&      317&       52\\\hline
Nevent$_{(b)}$            &      152&      100&       70&       36&        6\\\hline
$29sub/all$               &     0.86&     0.85&     0.85&     0.84&     0.82\\\hline
$R_{jet}$                 &     0.62&     0.62&     0.62&     0.62&     0.61\\\hline
                   Entries&    52542&    37741&    28477&    17189&     3142\\\hline
\hline
\end{tabular}
\vskip0.3cm
\caption{Selection 1. $\phigj=180^\circ \pm 15^\circ$. LUCELL algorithm.  ~$L_{int}=3 ~fb^{-1}$}
\vskip0.2cm
\begin{tabular}{||c||c|c|c|c|c|c|c|c|c|c||}
\hline
\hline
$Pt^{clust}_{CUT}$ &$\quad$  30 $\quad$&$\quad$  20 $\quad$&$\quad$ 15 $\quad$&$\quad$  10 $\quad$&$\quad\ $   5 $\quad\ $\\\hline
\hline
$Pt^{jet}$                &  212.521&  213.982&  214.736&  215.460&  216.044\\\hline
$\nuj$                    &    0.866&    0.850&    0.802&    0.742&    0.568\\\hline
$Pt_{(\nu)}^{Jet}$        &    0.869&    0.853&    0.805&    0.744&    0.569\\\hline
$R_{event}^{\nu\in Jet}$  &    0.038&    0.037&    0.036&    0.034&    0.028\\\hline
$Pt_{(\mu)}^{Jet}$        &    0.417&    0.390&    0.353&    0.336&    0.268\\\hline
$R_{event}^{\mu\in Jet}$  &    0.019&    0.019&    0.019&    0.017&    0.014\\\hline
$Pt^{miss}$               &    5.529&    5.487&    5.412&    5.337&    4.975\\\hline
$Pt^{miss}_{\nu\in Jet}$  &   24.076&   24.016&   23.622&   23.102&   21.347\\\hline
Nevent$_{(c)}$            &     1012&      694&      487&      261&       44\\\hline
Nevent$_{(b)}$            &      138&       90&       60&       27&        4\\\hline
$29sub/all$               &     0.86&     0.85&     0.84&     0.83&     0.80\\\hline
$R_{jet}$                 &     0.66&     0.65&     0.65&     0.65&     0.64\\\hline
                   Entries&    49253&    34775&    25582&    14562&     2786\\\hline
\hline
\end{tabular}
\end{center}
\end{table}


\begin{table}
\begin{center}
\large{ $300 < \Pt^{\gamma} < 360 ~GeV/c$}\\[20pt]
\normalsize
\caption{Selection 1. $\phigj=180^\circ \pm 15^\circ$. UA1 algorithm.  ~$L_{int}=3 ~fb^{-1}$}
\vskip0.2cm
\begin{tabular}{||c||c|c|c|c|c|c|c|c|c|c||}
\hline
\hline
$Pt^{clust}_{CUT}$ &$\quad$  30 $\quad$&$\quad$  20 $\quad$&$\quad$ 15 $\quad$&$\quad$  10 $\quad$&$\quad\ $   5 $\quad\ $\\\hline
\hline
$Pt^{jet}$                &  320.158&  321.502&  322.289&  322.869&  322.911\\\hline
$\nuj$                    &    1.077&    1.069&    1.060&    1.015&    1.161\\\hline
$Pt_{(\nu)}^{Jet}$        &    1.081&    1.072&    1.063&    1.018&    1.163\\\hline
$R_{event}^{\nu\in Jet}$  &    0.035&    0.034&    0.033&    0.033&    0.035\\\hline
$Pt_{(\mu)}^{Jet}$        &    0.515&    0.506&    0.476&    0.448&    0.433\\\hline
$R_{event}^{\mu\in Jet}$  &    0.019&    0.018&    0.017&    0.017&    0.017\\\hline
$Pt^{miss}$               &    5.764&    5.721&    5.692&    5.572&    5.691\\\hline
$Pt^{miss}_{\nu\in Jet}$  &   31.983&   32.597&   33.078&   32.371&   35.201\\\hline
Nevent$_{(c)}$            &      172&      117&       84&       46&        8\\\hline
Nevent$_{(b)}$            &       25&       15&       10&        6&        1\\\hline
$29sub/all$               &     0.84&     0.83&     0.82&     0.80&     0.78\\\hline
$R_{jet}$                 &     0.62&     0.62&     0.62&     0.62&     0.61\\\hline
                   Entries&    46297&    32513&    24157&    14318&     2642\\\hline
\hline
\end{tabular}
\vskip0.3cm
\caption{Selection 1. $\phigj=180^\circ \pm 15^\circ$. LUCELL algorithm.  ~$L_{int}=3 ~fb^{-1}$}
\vskip0.2cm
\begin{tabular}{||c||c|c|c|c|c|c|c|c|c|c||}
\hline
\hline
$Pt^{clust}_{CUT}$ &$\quad$  30 $\quad$&$\quad$  20 $\quad$&$\quad$ 15 $\quad$&$\quad$  10 $\quad$&$\quad\ $   5 $\quad\ $\\\hline
\hline
$Pt^{jet}$                &  320.687&  322.011&  322.732&  323.248&  323.646\\\hline
$\nuj$                    &    1.072&    1.061&    1.055&    0.983&    1.052\\\hline
$Pt_{(\nu)}^{Jet}$        &    1.076&    1.064&    1.057&    0.985&    1.053\\\hline
$R_{event}^{\nu\in Jet}$  &    0.035&    0.033&    0.033&    0.032&    0.032\\\hline
$Pt_{(\mu)}^{Jet}$        &    0.507&    0.483&    0.480&    0.412&    0.388\\\hline
$R_{event}^{\mu\in Jet}$  &    0.018&    0.018&    0.017&    0.016&    0.013\\\hline
$Pt^{miss}$               &    5.761&    5.722&    5.686&    5.499&    5.465\\\hline
$Pt^{miss}_{\nu\in Jet}$  &   32.220&   33.054&   33.221&   31.968&   34.611\\\hline
Nevent$_{(c)}$            &      161&      106&       74&       39&        7\\\hline
Nevent$_{(b)}$            &       22&       14&        9&        5&        1\\\hline
$29sub/all$               &     0.83&     0.82&     0.81&     0.80&     0.77\\\hline
$R_{jet}$                 &     0.66&     0.65&     0.65&     0.65&     0.64\\\hline
                   Entries&    43320&    29783&    21707&    12104&     2334\\\hline
\hline
\end{tabular}

\end{center}
\end{table}


\begin{thebibliography}{99}
\bibitem{1}
D0 Collaboration, F.~Abachi {\it et al.}, NIM {\bf A424} (1999)352.
\bibitem{2}
CDF Collaboration. F.~Abe {\it et al.}, Phys.Rev. {\bf D50} (1994)2966;
F.~Abe {\it et al.}, Phys.Rev.Lett. {\bf 73} (1994)225.
\bibitem{3}
D.~Denegri, R.~Kinnunen, A.~Nikitenko, CMS Note 1997/039 ``Study of
calorimeter calibration with $\tau$`s in CMS''.
\bibitem{4}
R.~Kinnunen, A.~Nikitenko, CMS Note 1997/097 ``Study of
calorimeter calibration with pions from jets in CMS''.
\bibitem{5}
J.~Womersley. A talk at CMS Week meeting, Aachen, 1997.
\bibitem{6}
J.~Freeman, W.~Wu, {\bf draft} ``In situ calibration of CMS HCAL calorimeter''.
\bibitem{7}
R.~Mehdiyev, I.~Vichou, ATLAS Note  ATL-COM-PHYS-99-054 (1999)
``Hadronic jet energy scale calibration using Z+jet events''.
\bibitem{8}
ATLAS Detector and Physics Performance, Technical Design Report, Volumes
{\bf 1, 2}, 1999. CERN/LHCC 99-14.
\bibitem{9}
N.B.~Skachkov, V.F.~Konoplyanikov D.V.~Bandourin,
``Photon -- jet events for calibration of HCAL''. Second Annual RDMS CMS Collaboration Meeting. CMS-Document, 1996--213. CERN, December 16-17, 1996, p.7-23.
\bibitem{10}
N.B.~Skachkov, V.F.~Konoplyanikov D.V.~Bandourin,
``$\gamma$-direct + 1 jet events for HCAL calibration''. Third Annual RDMS CMS Collaboration Meeting. CMS-Document, 1997--168. CERN, December 16-17, 1997,
p.139-153.
\bibitem{11}
T.~Sjostrand, Comp.Phys.Comm. {\bf 82} (1994)74.
\bibitem{12}
D.V.~Bandourin, V.F.~Konoplyanikov, N.B.~Skachkov.
``Jet energy scale setting with \gpj events at LHC
energies. Event rates, $\Pt$ structure of jet''.
JINR Communication, JINR, Dubna, 2000-
\bibitem{13}
D.V.~Bandourin, V.F.~Konoplyanikov, N.B.~Skachkov.
``Jet energy scale setting with \gpj events at LHC
energies. Minijets and cluster suppression and
$\Pt^{\gamma} - \Pt^{Jet}$ disbalance''.
JINR Communication, JINR, Dubna, 2000-
\bibitem{14}
D.V.~Bandourin, V.F.~Konoplyanikov, N.B.~Skachkov.
``Jet energy scale setting with \gpj events at LHC
energies. Selection of events with a clean \gpj topology and
$\Pt^{\gamma} - \Pt^{Jet}$ disbalance.''.
JINR Communication, JINR, Dubna, 2000-
\bibitem{15}
D.V.~Bandourin, V.F.~Konoplyanikov, N.B.~Skachkov.
``Jet energy scale setting with \gpj events at LHC
energies. Detailed study of the background suppression''.
JINR Communication, JINR, Dubna, 2000-
\bibitem{16}
S.~Abdullin, A.~Khanov, N.~Stepanov, CMS Note CMS TN/94--180 ``CMSJET''.
\bibitem{17}
 S.~Frixione, Phys.Lett. {\bf B429} (1998)369.
\bibitem{18}
S.~Catani, M.~Fontannaz and E.~Pilon, Phys.Rev. {\bf D58} (1998)094025


\end{thebibliography}
\end{document}